\newcommand\apjcls{1}
\newcommand\aastexcls{2}
\newcommand\othercls{3}

\newcommand\papercls{\aastexcls}
\documentclass[tighten,times,twocolumn]{aastex62}  

\if\papercls \apjcls
\usepackage{apjfonts}
\else\if\papercls \othercls
\usepackage{epsfig}
\usepackage{margin}
\usepackage{times}
\fi\fi
\usepackage{ifthen}
\usepackage{natbib}
\usepackage{bm}
\usepackage{amssymb, amsmath}
\usepackage{appendix}
\usepackage{etoolbox}
\usepackage[T1]{fontenc}
\usepackage{paralist}
\usepackage{newtxtext,newtxmath}
\if\papercls \apjcls
\newcommand\aas{\ref@jnl{AAS Meeting Abstracts}}
\newcommand\dps{\ref@jnl{AAS/DPS Meeting Abstracts}}
\newcommand\maps{\ref@jnl{MAPS}}
\else\if\papercls \othercls
\usepackage{astjnlabbrev-jh}
\fi\fi

\if\papercls \aastexcls
\hypersetup{citecolor=blue, 
            linkcolor=blue, 
            menucolor=blue, 
            urlcolor=blue}  
\else
\usepackage[
bookmarks=true,           
bookmarksnumbered=true,   
colorlinks=true,          
citecolor=blue,           
linkcolor=blue,           
menucolor=blue,           
urlcolor=blue,            
linkbordercolor={0 0 1},  
pdfborder={0 0 1},
frenchlinks=true]{hyperref}
\fi
\if\papercls \othercls

\else

\fi

\providecommand{\adsurl}[1]{\href{#1}{ADS}}

\makeatletter
\patchcmd{\NAT@citex}
  {\@citea\NAT@hyper@{%
     \NAT@nmfmt{\NAT@nm}%
     \hyper@natlinkbreak{\NAT@aysep\NAT@spacechar}{\@citeb\@extra@b@citeb}%
     \NAT@date}}
  {\@citea\NAT@nmfmt{\NAT@nm}%
   \NAT@aysep\NAT@spacechar\NAT@hyper@{\NAT@date}}{}{}

\patchcmd{\NAT@citex}
  {\@citea\NAT@hyper@{%
     \NAT@nmfmt{\NAT@nm}%
     \hyper@natlinkbreak{\NAT@spacechar\NAT@@open\if*#1*\else#1\NAT@spacechar\fi}%
       {\@citeb\@extra@b@citeb}%
     \NAT@date}}
  {\@citea\NAT@nmfmt{\NAT@nm}%
   \NAT@spacechar\NAT@@open\if*#1*\else#1\NAT@spacechar\fi\NAT@hyper@{\NAT@date}}
  {}{}
\makeatother

\makeatletter
\DeclareRobustCommand{\lowcase}[1]{\@lowcase#1\@nil}
\def\@lowcase#1\@nil{\if\relax#1\relax\else\MakeLowercase{#1}\fi}
\makeatother

\DeclareSymbolFont{UPM}{U}{eur}{m}{n}
\DeclareMathSymbol{\umu}{0}{UPM}{"16}
\let\oldumu=\umu
\renewcommand\umu{\ifmmode\oldumu\else\math{\oldumu}\fi}

\if\papercls \othercls

\else

\fi

\let\oldsim=\sim
\renewcommand\sim{\ifmmode\oldsim\else\math{\oldsim}\fi}
\let\oldpm=\pm
\renewcommand\pm{\ifmmode\oldpm\else\math{\oldpm}\fi}
\newcommand\by{\ifmmode\times\else\math{\times}\fi}

\newbox{\wdbox}
\renewcommand\c{\setbox\wdbox=\hbox{,}\hspace{\wd\wdbox}}
\renewcommand\i{\setbox\wdbox=\hbox{i}\hspace{\wd\wdbox}}

\newcount\timect
\newcount\hourct
\newcount\minct
\newcommand\now{\timect=\time \divide\timect by 60
         \hourct=\timect Cltiply\hourct by 60
         \minct=\time \advance\minct by -\hourct
         \number\timect:\ifnum \minct < 10 0\fi\number\minct}

\catcode`@=11

\newcommand\comment[1]{}

\newcommand\commenton{\catcode`\%=14}

\renewcommand\math[1]{$#1$}
\newcommand\mathshifton{\catcode`\$=3}

\let\atab=&
\newcommand\atabon{\catcode`\&=4}

\let\oldmsp=\sp
\let\oldmsb=\sb
\def\sp#1{\ifmmode
           \oldmsp{#1}%
         \else\strut\raise.85ex\hbox{\scriptsize #1}\fi}
\def\sb#1{\ifmmode
           \oldmsb{#1}%
         \else\strut\raise-.54ex\hbox{\scriptsize #1}\fi}
\newbox\@sp
\newbox\@sb
\def\sbp#1#2{\ifmmode%
           \oldmsb{#1}\oldmsp{#2}%
         \else
           \setbox\@sb=\hbox{\sb{#1}}%
           \setbox\@sp=\hbox{\sp{#2}}%
           \rlap{\copy\@sb}\copy\@sp
           \ifdim \wd\@sb >\wd\@sp
             \hskip -\wd\@sp \hskip \wd\@sb
           \fi
        \fi}
\def\msp#1{\ifmmode
           \oldmsp{#1}
         \else \math{\oldmsp{#1}}\fi}
\def\msb#1{\ifmmode
           \oldmsb{#1}
         \else \math{\oldmsb{#1}}\fi}

\def\supon{\catcode`\^=7}

\def\subon{\catcode`\_=8}

\def\supsubon{\supon \subon}

\newcommand\actcharon{\catcode`\~=13}

\newcommand\paramon{\catcode`\#=6}

\comment{And now to turn us totally on and off...}

\newcommand\reservedcharson{ \commenton  \mathshifton  \atabon  \supsubon 
                             \actcharon  \paramon}

\catcode`@=12
\reservedcharson

\if\papercls \apjcls

\else

\fi

\newcommand\chisq{\ifmmode{\chi\sp{2}}\else\math{\chi\sp{2}}\fi}
\newcommand\redchisq{\ifmmode{ \chi\sp{2}\sb{\rm red}}
                    \else\math{\chi\sp{2}\sb{\rm red}}\fi}
\newcommand\Teq{\ifmmode{T\sb{\rm eq}}\else$T$\sb{eq}\fi}
\newcommand\mjup{\ifmmode{M\sb{\rm Jup}}\else$M$\sb{Jup}\fi}
\newcommand\rjup{\ifmmode{R\sb{\rm Jup}}\else$R$\sb{Jup}\fi}
\newcommand\msun{\ifmmode{M\sb{\odot}}\else$M\sb{\odot}$\fi}
\newcommand\rsun{\ifmmode{R\sb{\odot}}\else$R\sb{\odot}$\fi}
\newcommand\mearth{\ifmmode{M\sb{\oplus}}\else$M\sb{\oplus}$\fi}
\newcommand\rearth{\ifmmode{R\sb{\oplus}}\else$R\sb{\oplus}$\fi}

\newcommand\Pran{\ensuremath{\mathrm{Pr}}}

\renewcommand{\bar}[1]{\overline{#1}}

\renewcommand{\bm}[1]{{\mbox{{\boldmath$#1$}}}}	

\shorttitle{Semiconvective layers in rapidly rotating flows. Consequences for giant planets}

\begin{document}


\title{Evolution of Semi-convective Staircases in Rotating Flows: Consequences for Fuzzy Cores in Giant Planets}


\author{J. R. Fuentes}
\affiliation{\rm Department of Applied Mathematics, University of Colorado Boulder, Boulder, CO 80309-0526, USA}

\author{Bradley W. Hindman}
\affiliation{\rm Department of Applied Mathematics, University of Colorado Boulder, Boulder, CO 80309-0526, USA}
\affiliation{\rm Department of Astrophysical and Planetary Sciences, University of Colorado, Boulder, CO 80309, USA}
\affiliation{\rm JILA, University of Colorado Boulder, Boulder, CO 80309-0440, USA}

\author{Adrian E. Fraser}
\affiliation{\rm Department of Applied Mathematics, University of Colorado Boulder, Boulder, CO 80309-0526, USA}

\author{Evan H. Anders}
\affiliation{\rm Kavli Institute for Theoretical Physics, University of California Santa Barbara, Santa Barbara, CA, 93106 USA}


\begin{abstract}
Recent observational constraints on the internal structure of Jupiter and Saturn suggest that these planets have ``fuzzy" cores, i.e., gradients of the concentration of heavy elements that might span a large fraction of the planet's radius. These cores could be composed of a semi-convective staircase, i.e., multiple convective layers separated by diffusive interfaces arising from double-diffusive instabilities. However, to date, no study has demonstrated how such staircases can avoid layer mergers and persist over evolutionary time scales. In fact, previous work has found that these mergers occur rapidly, leading to only a single convective layer. Using 3D simulations, we demonstrate that rotation prolongs the lifetime of a convective staircase by increasing the timescale for both layer merger and erosion of the interface between the final two layers. We present an analytic model for the erosion phase, predicting that rotation increases the erosion time by a factor of approximately $\mathrm{Ro}^{-1/2}$, where $\mathrm{Ro}$ is the Rossby number of the convective flows (the ratio of the rotation period to the convective turnover time). For Jovian conditions at early times after formation (when convection is vigorous enough to mix a large fraction of the planet), we find the erosion time to be roughly $10^{9}~\mathrm{yrs}$ in the non-rotating case and $10^{11}~\mathrm{yrs}$ in the rotating case. If these timescales are confirmed with a larger suite of numerical simulations, the existence of convective staircases within the deep interiors of giant planets is a strong possibility, and rotation could be an important factor in the preservation of their fuzzy cores.

\end{abstract}


\section{Introduction}

Determining whether gas giants are fully mixed is crucial for understanding their internal structure and evolution. The strength and distribution of compositional gradients affect heat transport, long-term cooling, and magnetism. Prior to the \textit{Juno} and \textit{Cassini} missions, gas giants were thought to be composed of a central core of heavy elements underneath a convective hydrogen-helium envelope. However, measurements of Jupiter's gravity field \citep{Bolton2017, Bolton2017b, Wahl2017, Militzer2022, Howard_2023} and seismology of Saturn's interior with waves in Saturn's rings \citep{Mankovich_2021} stand in conflict with that picture, and instead point toward ``fuzzy'' cores wherein heavy element abundances smoothly decrease from the core into the outer regions \citep[e.g.,][]{Helled2024}.

Recent formation models have shown that gas giants can form with composition gradients, but their subsequent survival for billions of years is still not well understood \citep{Helled2017,Muller2020,stevenson_et_al_2022}. This is because formation models predict that planets are born significantly hotter than today. Consequently, convection is so vigorous that the heavy element gradient everywhere except in the innermost region becomes mixed on short timescales $\sim $1 Myr \citep[e.g.,][]{Guillot_2004, Vazan2015, Muller2020}. 
Alternatively, a fuzzy core could possibly arise from a giant impact between a large planetary embryo and the proto-Jupiter/Saturn \citep{Liu2019}. However, the conditions required for such an impact (head-on collision, and $10\mearth$ impactor) are unlikely \citep{Helled2022}. 
\begin{figure*}
    \centering
    \includegraphics[width=\textwidth]{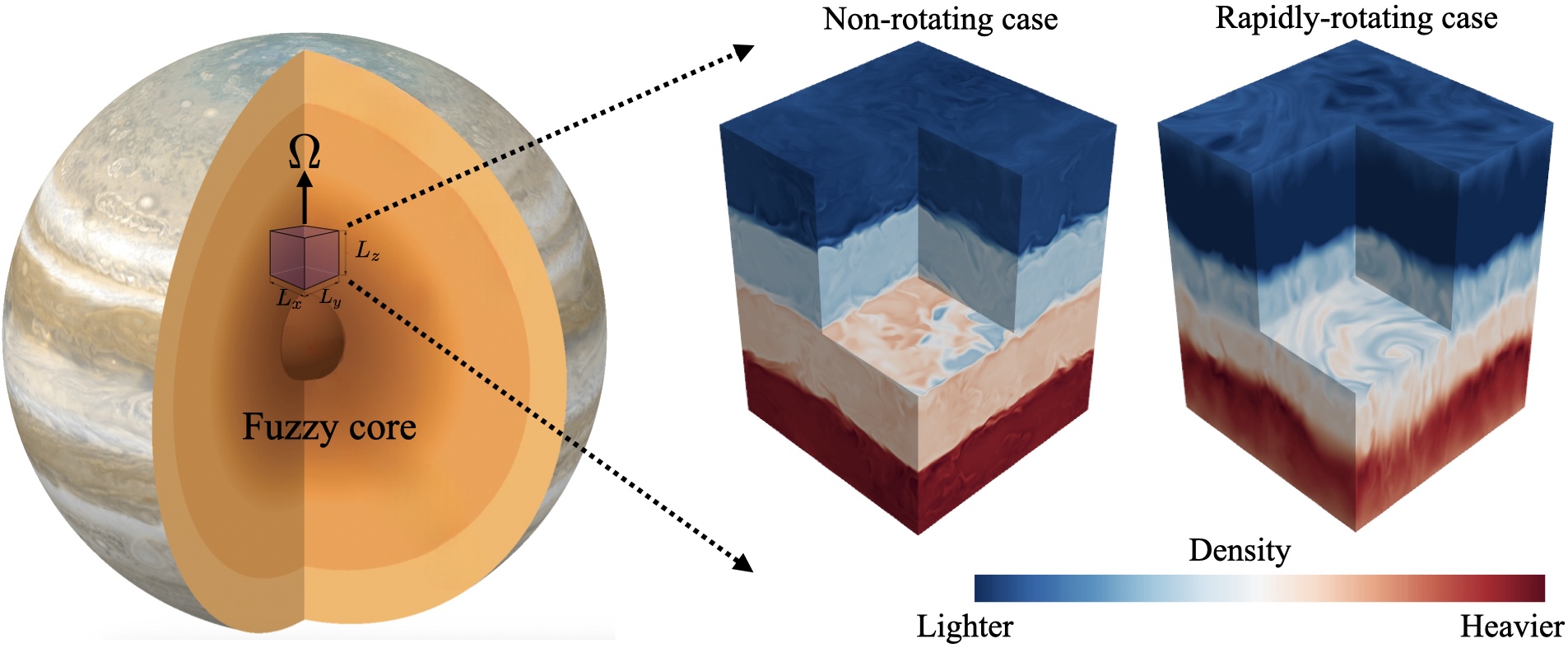}
    \caption{Left: Illustration of the model setup \citep[adapted from][with permission of the authors]{Ziv2024}. A local Cartesian domain is placed in the deep interior of the planet (not to scale). Right: 3D visualizations of the density field for the non-rotating case (left) and rotating case (right). For both simulations, the snapshots are shown at times when 4 convective layers are present in the fluid. As in experiments of pure thermal convection, the morphology of the flow is different depending on whether the flow is rotating or not. Both simulations have the same diffusivities, thermodynamic properties, and strength of thermal driving.}
    \label{fig:scheme}
\end{figure*}

While formation and evolution models, which are necessarily one-dimensional and spherically symmetric, have demonstrated that convection is efficient enough to mix most of the planet's interior, they lack a detailed treatment of how composition and heat are transported at convective boundaries. Further, by construction they do not capture instabilities that could prevent large-scale convection. In fact, under appropriate circumstances, the interaction of thermal and composition gradients, both present in the planet's interior, can trigger double-diffusive instabilities, leading to semiconvection \citep{Leconte_and_Chabrier_2012,Leconte_and_Chabrier_2013}. These hydrodynamical instabilities can form convective staircases, series of convective layers separated by sharp interfaces across which transport of heat and chemical species is achieved by molecular diffusion \citep[just like the many layers in a cafe latte which form due to the very same physics, see, e.g.,][]{Xue2017}. However, multi-dimensional simulations of semiconvection in astrophysical fluids have shown that staircases do not survive over long timescales as they have a tendency to merge quickly until a single well-mixed layer remains \citep[e.g.,][]{Mirouh2012,Wood2013,Moll2017,Zaussinger2019,Fuentes2022,Tulekeyev2024}. 

Rotation, another potential mechanism to mitigate mixing, has received less attention despite being a well-known factor that hampers thermal convection in oceanic, atmospheric, and stellar fluid dynamics \citep[e.g.,][]{Stevenson1979,Barker2014,Aurnou2020}. In our previous work \citep{Fuentes2023, Hindman2023}, we demonstrated that rotation significantly slows the advancement of an outer convection zone into a stably-stratified region by reducing the kinetic energy available for upward transport of heavy material. Given that rotation prevents strong mixing, it raises the question of whether rotation can also prevent the merger and prolong the lifetime of a convective staircase. 

In this work, we examine the effect of rotation on semiconvection, focusing our study on understanding the dynamical timescale of layer erosion. In Section~\ref{sec:numerics} we present and describe our numerical experiments. In Section~\ref{sec:erosion_model} we present an analytic model of the erosion of a convective-staircase that reproduces our numerical results. We conclude in Section~\ref{sec:discussion} with a summary and discussion of how our results
apply to gas giants.

\begin{figure*}
    \centering
    \includegraphics[width=\textwidth]{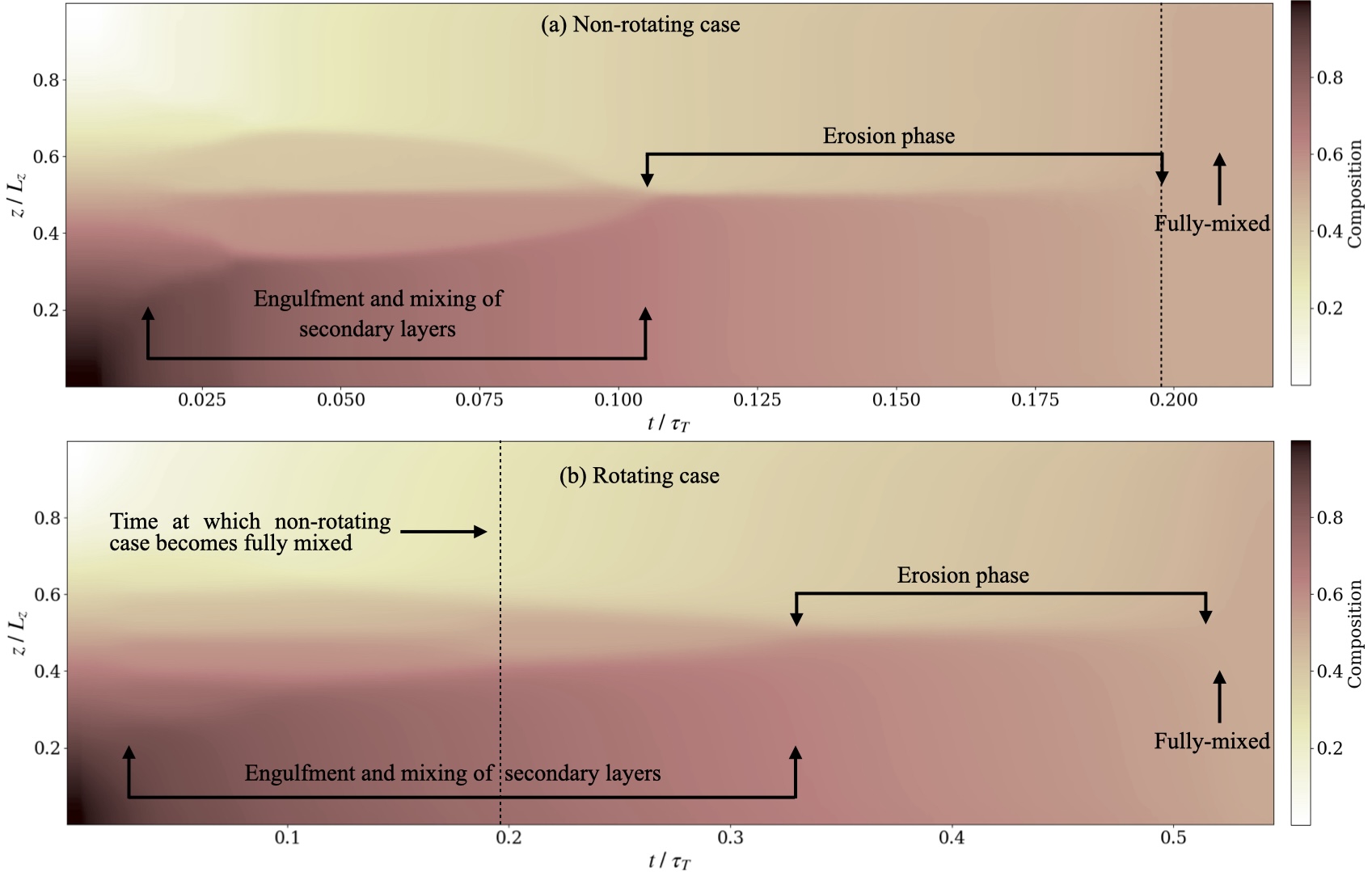}
    \caption{Horizontally-averaged concentration of heavy elements at depth $z$ and time $t$ (normalized by the thermal diffusion time across the box $\tau_{T}=L^2_z/\kappa_T$). Results are shown for both, the non-rotating and rotating cases (panels a and b respectively). Layer formation occurs in both cases, but layers persist for a longer time in the rotating case. The vertical dashed-line in both panels denotes the time at which the non-rotating case becomes fully mixed. Note that the horizontal axis is not the same in both panels.}
    \label{fig:evolution}
\end{figure*}
\section{Numerical simulations}\label{sec:numerics}

In the following, we present two hydrodynamic simulations, with and without rotation, to highlight the effects of rotation on the evolution of a convective staircase. We initialize the fluid so that both the temperature and the concentration of heavy elements (and thus the mean molecular weight) increase linearly with depth. The mean molecular weight gradient is strong enough relative to the temperature gradient such that the combined density increases with depth (i.e., the system is stable against over-turning convection according to the Ledoux criterion). This configuration is susceptible to a double-diffusive instability. After an initial phase of weak, instability-driven turbulence, layer formation occurs due to the interaction of the turbulent fluxes and background gradients of heat and mean molecular weight  \citep[see the excellent review by][]{Garaud2021}.

For simplicity, we focus on a small polar region deep within the planet's interior (see Figure~\ref{fig:scheme}), such that the depth of the domain $L_z$ is much smaller than a density scale height (comparable to the planet's radius) and the flows are subsonic. Under those conditions, the Boussinesq approximation is valid, and the effects of curvature can be neglected \citep{Spiegel_Veronis_1960}. We model this region using a Cartesian domain ($x$, $y$, $z$) with constant gravity and uniform rotation, both aligned with the vertical axis: $\bm{g} = -g\bm{\hat{z}}$, $\bm{\Omega} = \Omega\bm{\hat{z}}$. The density of the fluid depends on both composition and temperature.
In what follows, all results are presented in dimensionless form. For additional details on the numerical setup, nondimensionalization, and code, we refer the reader to the Appendix.

As seen in snapshots of the simulations (Figure~\ref{fig:scheme}), rotation significantly affects the spatial scales of the convective flow. In the non-rotating case, the length scales are approximately isotropic, while in the rotating case, the horizontal length scale of the flow is much smaller than the vertical scale \citep[with vortices and thin columns, as expected from the Taylor-Proudman theorem, e.g.,][]{Proudman1916,Taylor1917}. Note for the run without rotation, the density within each individual convection layer is approximately uniform, whereas in the rotating case, each convective layer has a small residual density gradient (see also Figure~\ref{fig:density_evolution}). Even when composition is well-mixed, the inefficiency of the vertical heat transport in rotating convection is well known to develop a small vertical temperature gradient in the layer \citep[see, e.g.,][]{Julien1996,Julien1996_1}.

\subsection{Qualitative evolution of a convective staircase}\label{sec:results}

Figure~\ref{fig:evolution} provides a summary of the evolution of the concentration of heavy elements present in the simulation. The color scale indicates the horizontal average of the compositional concentration $C$ as a function of nondimensional time, $t/\tau_T$, and height within the fluid layer, $z/L_z$.  Time is measured in thermal diffusion times defined by $\tau_{T} = L_z^2/\kappa_T$, where $\kappa_T$ is the thermal diffusivity. Initially, the composition has a linear distribution with height, but over a relatively short amount of time ($\approx$ 0.017 and 0.025 for the nonrotating and rotating simulations, respectively) the fluid develops multiple well-mixed convective layers separated by stably-stratified interfaces. We find that both  the nonrotating and rotating models form 6 layers (see Figure~\ref{fig:evolution}). Once the layers form, they immediately begin to merge.

For example, in the nonrotating model (Figure~\ref{fig:evolution}a), the two innermost and the two outermost layers grow in thickness and engulf the layers sandwiched in between. This process completes around $t\approx 0.03~\tau_T$, leaving four layers, two inner layers and two outer layers. The outer layers then start to devour the inner layers with the engulfment requiring $\sim 0.07$ more thermal diffusion times.  The penultimate state, two layers of roughly equal size, is achieved around $t\approx 0.1~\tau_{T}$. These two final layers persist over time, since there is a density jump across the common interface that prevents their instantaneous merger and mixing. However, the layers slowly mix across the interface and the density difference between them decreases until the final merger at roughly $t=0.2~\tau_{T}$. Hence, the erosion of the final interface takes roughly one-tenth of a thermal diffusion time. The rotating model (Figure~\ref{fig:evolution}b), appears to undergo similar processes, except that the mergers and interface erosion take much longer to occur. The initial layer mergers do not complete until $t=0.33~\tau_{T}$ and the final internal interface takes roughly 0.2 thermal diffusion times to erode, i.e., the merging of the initial layers and the erosion of the final two layers take approximately three times and two times longer, respectively, than in the non-rotating case.

\begin{figure}
    \centering
    \includegraphics[width=\columnwidth]{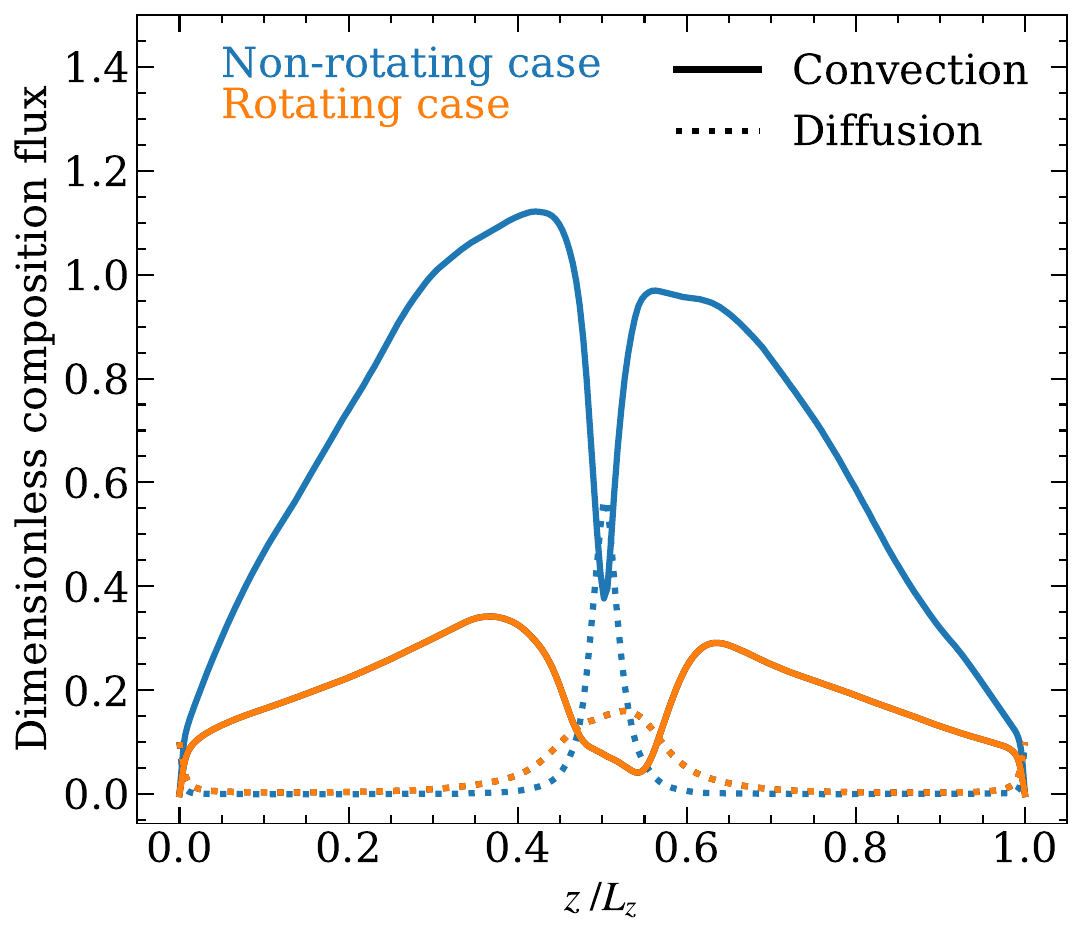}
    \caption{Vertical profiles of the convective and diffusive composition fluxes for the non-rotating case (blue) and rotating case (orange). The fluxes are shown at the beginning of the erosion phase of the remaining two layers. The solid and dashed lines denote the contribution to the flux purely due to convection and diffusion, respectively.}
    \label{fig:fluxes}
\end{figure}

We emphasize that the erosion process is a turbulent mixing mechanism (i.e., entrainment) not a diffusive one. As show in Figure~\ref{fig:fluxes}, the compositional transport across the interface is mostly dominated by the convective flux in both the rotating and the non-rotating simulations. Certainly, in the outer portions of the interface, turbulent mixing dominates. Even within the very center of the interface, diffusion only carries roughly 50--60\% of the composition flux with the remainder carried by turbulent mixing. So while, diffusion does play a role in chemical transport across the interface, it does so primarily by controlling the width of the interface. As shown by \cite{Fernando1989}, in a two-layer system, the equilibrium thickness of the diffusive interface is governed by a balance between thickening due to diffusion and thinning caused by convective entrainment. It is also worth mentioning that Under Jovian conditions, the diffusivity ratio is smaller by an order of magnitude with respect to our simulations (i.e. $\kappa_C/\kappa_T \sim 0.01$). Adopting a smaller chemical diffusivity would further diminish the role of compositional diffusion leading to its subdominance in interface erosion.

\begin{figure}
    \centering
    \includegraphics[width=\columnwidth]{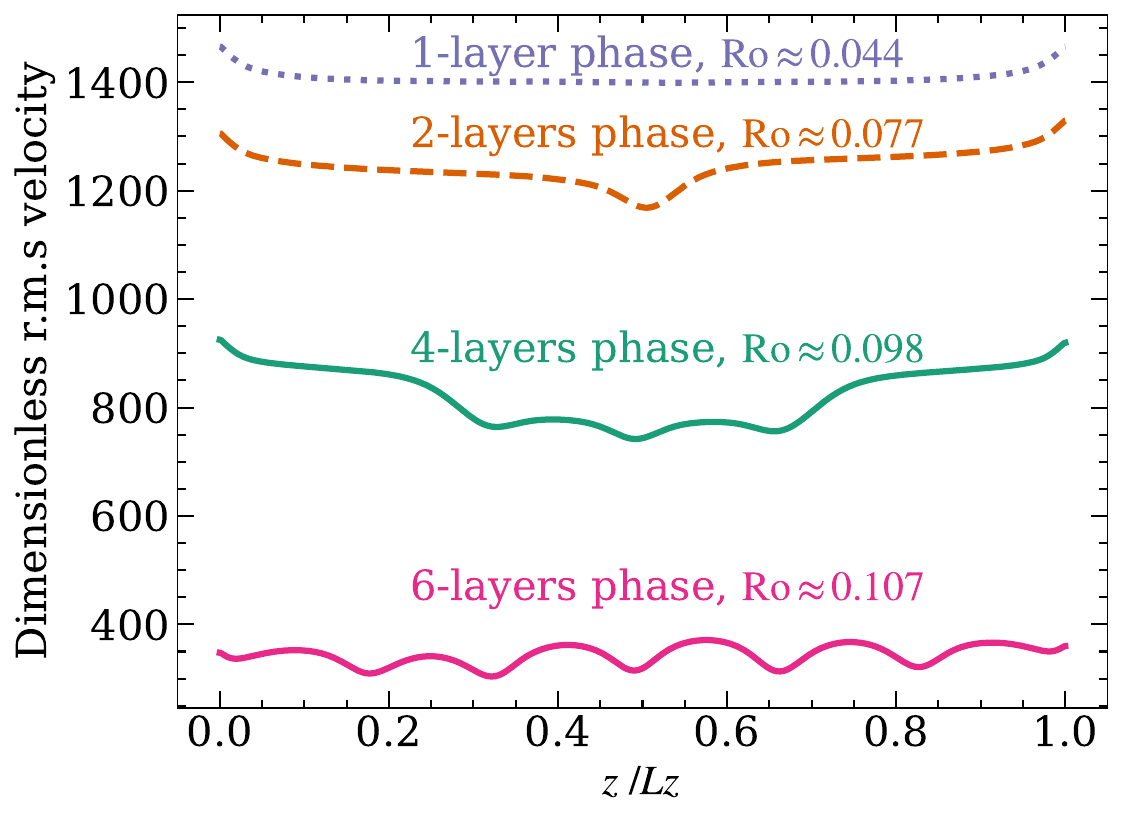}
    \caption{Vertical profiles of the r.m.s velocity for 4 different phases during the evolution of the convective staircase for the rotating case. From the r.m.s velocity, we have estimated the Rossby number, $\mathrm{Ro}$, of the flow (see main text). As expected, as layers merge the flow velocity increases. Time increases as the system evolves from 6 to 4 to 2 and finally to 1 layer.}
    \label{fig:velocity}
\end{figure}

As we show later, an important parameter for the mixing timescales of the fluid is the Rossby number, defined as the ratio of the rotational period to the convective turnover time, $\mathrm{Ro}\equiv U_{\mathrm{conv}}/2\Omega \ell$, where $U_{\mathrm{conv}}$ is the characteristic scale of the convective velocity, and $\ell$ is the characteristic lengthscale of the convection. Rapidly rotating flows have $\mathrm{Ro}\ll 1$, and the efficiency of convective mixing is much smaller than for non-rotating flows. In our simulations, we calculate Ro using the r.m.s. velocity of the fluid in each convective layer as a proxy for $U_{\mathrm{conv}}$, and set $\ell$ to the thickness of a convective layer in the staircase. Figure~\ref{fig:velocity} shows profiles of the r.m.s. velocity for the rotating case, at different stages during the evolution of the staircase. As expected, we find that as layers merge, the freefall distance of each surviving layer increases, leading to an increase in the convective velocity (see Equation~\ref{eq:u_conv} for how $U_{\mathrm{conv}}$ depends on $\ell$). However, the ratio $U_{\mathrm{conv}}(\ell)/\ell$ decreases. Consequently, the Rossby number, shown at the top of each curve in Figure~\ref{fig:velocity}, decreases after each merger, starting with $\mathrm{Ro} \approx 0.107$ in the 6-layers phase, and ultimately reaching $\mathrm{Ro}\approx 0.044$. We also note that initially, when there are 6 layers in the fluid, the velocity and size of each layer is uniform. Later on, the outer convective layers have a larger speed than the interior layers, engulfing them and leaving only two layers that slowly erode each other until a single convective layer remains.

\section{Analytic model for the erosion phase}\label{sec:erosion_model}

A complete characterization of all the timescales involved in the problem, and how they change with the choice of parameters is beyond the scope of this work. However, in what follows, we present a detailed model for the erosion stage that reproduces our numerical results.
\begin{figure}
    \centering
    \includegraphics[width=\columnwidth]{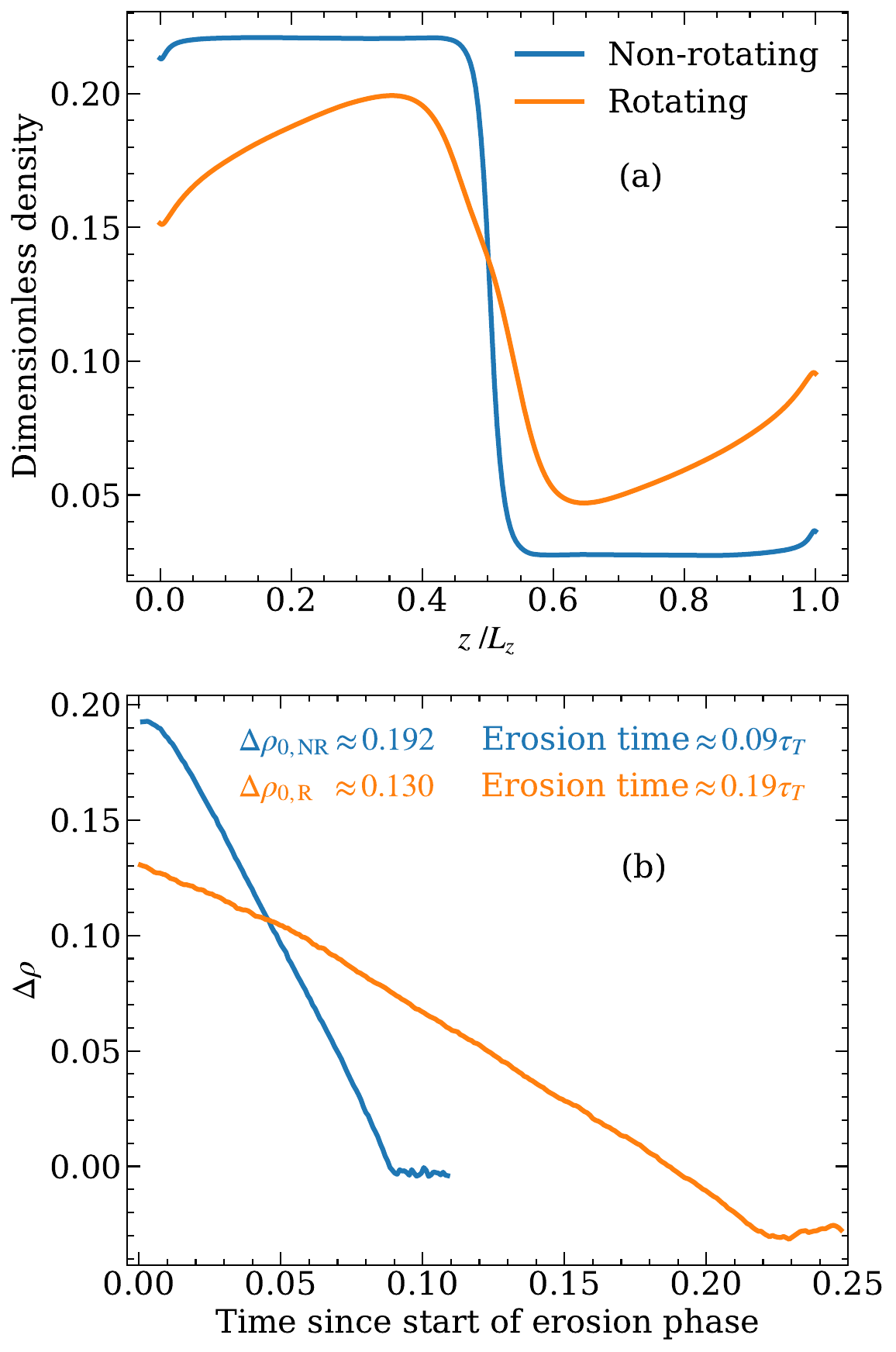}
    \caption{Panel (a): Density profile at the beginning of the erosion phase for the non-rotating and rotating cases. Note that in the rotating case, the density is not constant within the convection zone. Instead, there is a positive density gradient due to the thermal structure of the fluid in the convection zone (in nonrotating convection, the temperature becomes isothermal whereas in rotating convection, the fluid develops a small temperature gradient). Panel (b): Evolution of the density contrast between the two layers during the erosion phase. For both cases $\Delta\rho$ decreases linearly with time, but the rotating case decays at a slower rate. Further, the initial density contrast at the beginning of the erosion phase is smaller for the rotating case (likely due to diffusion since it takes longer for the fluid to reach this stage).}
    \label{fig:density_evolution}
\end{figure}

When the layered system has reached its penultimate state, it has two well-mixed, convecting layers of equal thickness ($L_z/2$) separated by a thin stably stratified interface (see  Figure~\ref{fig:density_evolution}a). The mass density of the fluid in the upper layer is less than the density of the fluid in the lower layer. The convection in each layer is maintained by a constant heat flux $F_H$ driven through the system. Due to the partial mixing that occurs across the thin interface, the density contrast between the two layers, $\Delta\rho$, decreases with time (see Figure~\ref{fig:density_evolution}b), changing the gravitational potential energy across the layer at a rate $d(\Delta E)/dt = - (1/8) g L^2_z d(\Delta\rho)/dt$. If we assume that the energy required for mixing across the interface comes from and is proportional to the kinetic energy flux in each convective layer \citep[``the entrainment hypothesis'', e.g.,][]{Linden1975}, one can derive an evolution equation for the density difference between the two layers,

\begin{equation}\label{eq:drho}
\dfrac{d\Delta E}{dt} = -\dfrac{g L^2_z}{8} \dfrac{d\Delta\rho}{dt} \approx 2\times \frac{\varepsilon}{2}\rho_0 U_{\mathrm{conv}}^3 \; ,
\end{equation}
where the preceding factor of 2 in the final expression accounts for entrainment from above and below the interface, $\rho_0$ is the average density of the fluid, and $\varepsilon$ is the mixing efficiency \citep[$\approx 0.5$--$1$, e.g.,][]{Fuentes2020}.

We see that given $U_{\mathrm{conv}}$, Equation~\eqref{eq:drho} can be integrated to obtain $\Delta\rho(t)$. In the non-rotating and rapidly-rotating limits, typical convective speeds ($U_{\rm NR}$ and $U_{\rm R}$, respectively) can be estimated using mixing-length theory and turbulent, diffusion-free scaling laws for convective heat transfer \citep[e.g.,][] {Stevenson1979, Ingersoll1982, Barker2014, Aurnou2020}. Following the derivation of \cite{Fuentes2023}, we obtain

\begin{equation}
U_{\mathrm{NR}}\sim \left(\dfrac{\alpha g F_H \ell}{\rho_0 c_P}\right)^{1/3}\,,\hspace{0.25cm} U_{\mathrm{R}}
\sim \left(\dfrac{\alpha g F_H}{\rho_0 c_P}\right)^{2/5}\left(\dfrac{\ell}{2\Omega}\right)^{1/5}~.\label{eq:u_conv}
\end{equation}
In these expressions, $\alpha$ is the coefficient of thermal expansion, $c_P$ is the specific heat capacity at constant pressure, and $\ell = L_z/2$ is the approximate thickness of each convective layer. Since these speeds are temporally constant, integration of Equation~\eqref{eq:drho} gives

\begin{equation}\label{eq:delta_rho}
\Delta\rho(t) = \Delta\rho_0\left(1-\dfrac{t}{\tau}\right),
\end{equation}
where $\Delta \rho_0$ is the initial density contrast ($t=0$ measured from the beginning of the erosion phase), and $\tau$ is the time required for $\Delta\rho \rightarrow 0$ (the mixing timescale of the two convective layers). Assuming that $\varepsilon \approx 1$ for both cases, but taking into account that $\Delta\rho_0$ is different, we find

\begin{equation}
\tau_{\mathrm{NR}}\sim \dfrac{\Delta\rho_{0,\mathrm{NR}} c_P L_z}{4\alpha F_H} \,,\hspace{0.25cm} \tau_{\mathrm{R}}
\sim~\tau_{\mathrm{NR}} \left(\dfrac{\Delta\rho_{0,\rm{R}}}{\Delta\rho_{0,\rm{NR}}}\right)\mathrm{Ro}^{-1/2}~,\label{eq:T_erosion}
\end{equation}
where $\mathrm{Ro} = U_R/2\Omega \ell$ is the Rossby number of the flow (approximately the same in each convective layer), defined in terms of the rotationally-constrained convective velocity. Therefore, in a rapidly rotating system where $\mathrm{Ro}\ll 1$, the erosion timescale of the convective interface increases by a factor of $\mathrm{Ro}^{-1/2}$ (the ratio between the initial density jumps is of order unity). This is the same relation that we found in our previous paper for the timescale for an outer convection zone to mix a layer with a stable compositional gradient \citep[see Equation~10 in][]{Fuentes2023}.

We find a good agreement between the model and our simulations. First, the decrease in the density jump across the interface is linear in time (see Figure~\ref{fig:density_evolution}b), just as predicted by Equation~\eqref{eq:delta_rho}. Second, when comparing the mixing timescales, the ratio between the erosion time in the rotating case and the non-rotating case is $(0.19/0.09)\approx 2.11$. Equation~\eqref{eq:T_erosion} predicts that this ratio is given by $(\Delta\rho_{0,\rm{R}}/\Delta\rho_{0,\rm{NR}})\mathrm{Ro}^{-1/2}$. In our numerical results, we find $\mathrm{Ro}\approx 0.077$ in the erosion phase (see Figure~\ref{fig:velocity}), and $(\Delta\rho_{0,\rm{R}}/\Delta\rho_{0,\rm{NR}}) \approx 0.130/0.192 \approx 0.68$ (Figure~\ref{fig:density_evolution}b). Then, the predicted ratio is $\approx 0.68\times (0.077)^{-1/2}\approx 2.44$, just like in the simulations. The small (but negligible) differences between the numerical results and the model could be due to the effects of thermal diffusion, which we did not take into account, and possibly a mixing efficiency that depends weakly on rotation. 

Given the agreement between the model and our numerical results, it is encouraging to compute values of the timescales relevant for gas giants. For typical Jovian conditions in the deep interior, at $t\sim 10^{8}~\mathrm{yrs}$ after formation (where a large fraction of the planet is expected to be convective)

\begin{align}
\nonumber \tau_{\mathrm{NR, J}}\sim 10^9~\mathrm{yrs} &\left(\dfrac{\Delta \rho_0}{1~\rm g~cm^{-3}}\right) \left(\dfrac{c_P}{10^8~\mathrm{erg~g^{-1}~K^{-1}}}\right) \left(\dfrac{L_z}{10^9~\mathrm{cm}}\right)\\
&\times  \left(\dfrac{\alpha}{10^{-5}~\rm K^{-1}}\right)^{-1}\left(\dfrac{F_H}{ 10^5~\mathrm{erg~cm^{-2}~s^{-1}}}\right)^{-1}~,
\end{align}
and

\begin{align}
\tau_{\mathrm{R, J}}\sim 10^{11}~\mathrm{yrs}\left(\dfrac{\tau_{\mathrm{NR,J}}}{10^9~\mathrm{yrs}}\right)\left(\dfrac{\mathrm{Ro}}{10^{-5}}\right)^{-1/2}~,
\end{align}
where the numbers were inferred from \citet{Cumming2018} and \citet{Helled2022}. This is a difference of two orders of magnitude, and the mixing time is longer than the current age of Jupiter or Saturn. We emphasize that evaluating the numbers at later times would give even longer timescales, since convection becomes weaker as the planet's luminosity decreases with time.

\section{Discussion}\label{sec:discussion}
 Our results suggest that a convective staircase can persist over long timescales provided that convection is rotationally-constrained ($\mathrm{Ro}\ll 1$), a condition that is satisfied by giant planets. This is because rapid rotation changes the morphology of the convective flow (dominated by thin plumes) and reduces the kinetic energy flux available for compositional mixing, as recently shown by \cite{Fuentes2023}.

Despite the simplifications made in our model (e.g., ignoring spherical geometry and density stratification) and the uncertainties in the interior parameters of gas giants (particularly for $\Delta\rho_0$ and the mixing efficiency $\varepsilon$), we have demonstrated that the effect of rotation is substantial in rapidly rotating gas giants, increasing the erosion time of the staircase by two orders of magnitude, for Jovian conditions at $t\sim 10^8~\mathrm{yrs}$ after formation. The effect is even larger at later times since the convective strength becomes smaller as the planet's luminosity decreases with time. Also, a smaller value of the mixing efficiency for the rotating case will just make the timescale longer, once again supporting our hypothesis that rotation prolongs the lifetime of a convective staircase.

Characterizing the timescale of the initial mergers is also of great interest, as it would allow us to determine the current evolutionary stage of the fuzzy cores in Jupiter and Saturn. For example, Saturn’s seismology indicates the presence of a stably stratified region, which likely requires a semi-convective staircase with many steps, giving the planet a dense spectrum of gravity modes. If Saturn were in the erosion phase, there would be only a single interface and hence a single interface mode in the pulsation spectrum, which contradicts the ring-seismology inferences \citep{Fuller2014,Belyaev2015,Mankovich_2021}.
Unlike the erosion phase, where the staircase consists of only two steps of equal size throughout the process, the initial mergers occur in a fluid region with many steps of size $h$, which merge over time, causing $h$ to increase. From Equations (6) and (7) in \cite{Fuentes2023}, and using the same parameters, the growth rate of a convective layer by engulfing fluid from a stable region below is $h_{\mathrm{NR}}\sim 3 \times 10^9~\mathrm{cm}~(t/4.5~\mathrm{Gyr})^{1/2}$ for the non-rotating case, and $h_{\mathrm{R}}\sim 2\times 10^8~\mathrm{cm}~(t/4.5~\mathrm{Gyr})^{5/12}$ for the rotating case. This suggests that over the age of the solar system, the planet would be nearly fully mixed in the non-rotating case, while about $\sim 10$ convective layers would remain in the staircase for the rotating case.

Another factor that could influence convective mixing and the lifespan of a staircase is the presence of magnetic fields. 
For example, the preference for fluid flows to move parallel to the magnetic field direction means that a vertical field geometry can significantly reduce horizontal mixing when the field strength is sufficiently large \citep[e.g.,][]{Yan2019}. Recently, \cite{Sangui2022} investigated semiconvection in the presence of a uniform vertical background magnetic field, finding that a sufficiently strong field delays layer formation and substantially reduces the internal flux across the layers. However, different field geometries could lead to different results \citep[e.g.,][]{Calkins2023}.

We encourage future work to either confirm or refute our findings using a larger suite of numerical simulations. More detailed models of the growth and merging rates of semiconvective layers would significantly improve the estimates mentioned above and help assess the possible structures of giant planets over evolutionary timescales. It is also important to emphasize that semiconvection is not only relevant for the thermal evolution and transport processes in giant planets, but also in stars \citep[e.g.,][]{Spruit2013,Zaussinger2013,Ding2014,Moore2016}. Recently, semiconvection has been shown to be crucial for understanding the internal dynamics, magnetism, and potential habitability of icy moons \citep[e.g.,][]{Vance2021,Wong2022,Naseem2023,Idini2024}. The Rossby number in icy moons is also small \citep[$\mathrm{Ro}\ll 1$, see, e.g.,][]{Bire2022} and therefore the results obtained in this work could be applicable to them.


\begin{acknowledgements}
We thank Keith Julien for the countless conversations on fluid dynamics, in particular the ones on rotationally-constrained turbulence. You were not only a great scientist, but also a good friend, with an infectious smile.  We also thank Pascale Garaud and Jim Fuller for useful conversations on double-diffusive convection and about the application of our results to Saturn and Jupiter. J.R.F. is supported by NASA through the Solar System Workings grant 80NSSC24K0927, and Heliophysics grants 80NSSC19K0267 and 80NSSC20K0193. B.W.H. is supported by NASA Heliophysics through grants 80NSSC20K0193, 80NSSC24K0125, and 80NSSC24K0271. A.F. acknowledges support from NASA HTMS grant 80NSSC20K1280, and from the George Ellery Hale Postdoctoral Fellowship in Solar, Stellar and Space Physics at CU Boulder. E.H.A. acknowledges support from NSF grant PHY-2309135 and Simons Foundation grant (216179, LB) to the Kavli Institute for Theoretical Physics (KITP).
\end{acknowledgements}

\appendix \label{sec:appendix}

\section{Fluid equations and numerical methods}

We express the fluid quantities as the sum of a linear hydrostatic background (denoted by the subscript 0) and a dynamic perturbation to the background (denoted by primes), e.g., the total temperature and composition are expressed as $T = T_0(z) + T'$, and $C = C_0(z) + C'$, respectively. The density perturbations satisfy $\rho'/\rho_0 \ll 1$, and are related to $T'$ and $C'$ through $\rho' = \rho_0(\beta C' - \alpha T')$, as demanded by the Boussinesq approximation \citep{Spiegel_Veronis_1960}. Here, $\beta$ and $\alpha$ are the coefficients of compositional and thermal contraction/expansion (both assumed positive constants), respectively.
Before presenting the fluid equations, we non-dimensionalize them using $[T] = |\partial_z T_0| L_z$, $[C] = |\partial_z C_0| L_z$ as units of temperature and composition. We use the domain's depth $L_z$ as the unit of length, and the thermal diffusion time $\tau_T = L^2_z/\kappa_T$ (where $\kappa_T$ is the thermal diffusivity) as the unit of time. By this choice, a unit of pressure corresponds to $[P] = \rho_0 (\kappa_T/L_z)^2$. The dimensionless equations are 

\begin{align}
&\nabla \cdot \bm{u} = 0\, ,\\
&\dfrac{D\bm{u}}{Dt} + \dfrac{\rm Pr}{\rm Ek} \bm{\hat{z}}\times \bm{u}  = - \nabla P' + \mathrm{Ra}\mathrm{Pr}\left(R_\rho C' - T'\right)\bm{\hat{z}} + \mathrm{Pr} \nabla^2\bm{u} \, ,\\
&\dfrac{D C}{Dt} = \mathrm{Le}^{-1} \nabla^2 C\, ,\\
&\dfrac{D T}{D t} = \nabla^2 T\, ,
\end{align}
where $D/Dt = \partial_t + \bm{u}\cdot \bm{\nabla}$, and $\bm{u}$ is the velocity field.

There are 5 dimensionless numbers that characterize the evolution of the flow. These are the Rayleigh number Ra, Ekman number Ek, density ratio $R_\rho$, Prandtl number Pr, and Lewis number Le, which are defined respectively as
\begin{align}
    &\mathrm{Ra} = \frac{\alpha g |\partial_z T_0|L^4_z}{\kappa_T \nu} ,\hspace{0.1cm}
    \mathrm{Ek} = \frac{\nu}{2\Omega L^2_z},\\
   & R_\rho = \frac{\beta |\partial_z C_0|}{\alpha |\partial_z T_0|},\hspace{0.1cm}
    \Pran = \frac{\nu}{\kappa_T},\hspace{0.1cm}
    \mathrm{Le} = \frac{\kappa_T}{\kappa_C},
\end{align}
where $\Omega$ is the rotation rate, $\nu$ is the kinematic viscosity, and $\kappa_T$ and $\kappa_C$ are the thermal and compositional diffusivities, respectively. $R_\rho$ is the ratio of the stabilizing and destabilizing effect of the compositional and thermal buoyancy.

We initialize the fluid with linear distributions of temperature and composition,  $T_0(z) = 1-z, \quad C_0(z) = 1-z$, and fix the flux of temperature and composition at the top and bottom of the domain, so the boundary conditions for temperature and composition are $\partial_z T|_{z=0,1} = \partial_z C|_{z=0,1} = -1$~.
The boundary conditions for the velocity are impenetrable and stress free at both boundaries ($\hat{z}\cdot\bm{u} = \hat{x}\cdot\partial_z\bm{u} = \hat{y}\cdot\partial_z\bm{u} = 0$ at $z = 0,1$).

The simulations in this work use $\mathrm{Ra} = 2\times 10^{9}$, $\mathrm{Pr} = \mathrm{Le}^{-1} = 0.1$, and $\mathrm{Ek} = 3\times 10^{-6}$. We set the Coriolis force to be identically zero in the non-rotating case ($\mathrm{Ek} = \infty$). Double-diffusive instabilitites are expected to occur when the density ratio lies within a finite range $R_\rho \in [1,~ (\mathrm{Pr} + 1)/(\mathrm{Pr} + \mathrm{Le}^{-1})]$ \citep[e.g.,][]{Garaud2018,Garaud2021}. However, it is likely that layer formation occurs only in a narrower range $R_\rho \in [1, ~ R_{\mathrm{max}}]$, where $R_\mathrm{max}$ is uncertain but has been estimated to be $R_{\mathrm{max}}\approx \mathrm{Pr}^{-1/2}$ \citep{Mirouh2012,Wood2013}. For $R_\rho > R_{\mathrm{max}}$, the fluid evolves into a state of enhanced diffusion instead of forming layers \citep{Mirouh2012,Wood2013}. In this work we have $\mathrm{Pr} = {\rm Le}^{-1} =0.1$, giving the upper bound for double-diffusive instabilities as $(\mathrm{Pr} + 1)/(\mathrm{Pr} + \mathrm{Le}^{-1}) \approx 5.5$, and $R_{\mathrm{max}} \approx 3.2$. 
\cite{Moll2017_rot} showed that the onset of double-diffusive instabilities is not directly affected by rotation, but it is likely that $R_{\mathrm{max}}$ depends on the rotation rate. For simplicity, and to ensure layer formation in all our runs, we set the density ratio to $R_\rho = 1.25$.

Under Jovian conditions, $\mathrm{Ra} = \mathcal{O}(10^{43})$, $\mathrm{Ek} = \mathcal{O}(10^{-17})$, and $\mathrm{Le}^{-1} \approx \mathrm{Pr} = \mathcal{O}(10^{-2})$. Unfortunately, the value of the density ratio $R_\rho$ is unknown given our ignorance on the compositional and thermal stratification in the deep interiors of the gas giants \citep[see, e.g.,][]{Guillot_2004,French2012}. However, by fixing $\mathrm{Ra}\gg 1$, $\mathrm{Ek} \ll 1$, $\mathrm{Pr} < 1$, and $\mathrm{Le^{-1} < 1}$, we ensure that our simulations are qualitatively in the same dynamical regime as gas giants.

We time-evolve equations A1--A4 using the Dedalus pseudospectral solver \citep{Burns2020} version 3, using timestepper SBDF2 \citep{wang_ruuth_2008} and CFL safety factor 0.2. All variables are represented using a Chebyshev series with 384 terms for $z \in [0, 1]$ and Fourier series with 384 terms in the periodic $x$ and $y$ directions, where $x,~y \in [0, 0.75]$. We use the 3/2-dealiasing rule in all directions, so that nonlinearities are calculated in physical space on a $576^3$ grid.
To start our simulations, we add random-noise temperature perturbations sampled from a normal distribution with a magnitude of $10^{-5}$ compared to the initial temperature field.  Using  2048 cores, the non-rotating and rotating runs took 20 and 55 days to finish, respectively.

\bibliography{references}{}
\bibliographystyle{aasjournal}

\end{document}